\documentclass{article}

\usepackage{arxiv}

\usepackage[utf8]{inputenc} 
\usepackage[T1]{fontenc}    
\usepackage{url}            
\usepackage{booktabs}       
\usepackage{amsfonts}       
\usepackage{nicefrac}       
\usepackage{microtype}      
\usepackage{lipsum}		
\usepackage{graphicx}
\usepackage{natbib}
\usepackage{doi}

\usepackage{graphicx}
\usepackage{multicol,multirow}
\usepackage{amsmath,amssymb,amsfonts}
\usepackage{mathrsfs}
\usepackage{mathalpha}
\usepackage{amsthm}
\usepackage{rotating}
\usepackage{appendix}
\usepackage{ifpdf}
\usepackage[T1]{fontenc}
\usepackage{newtxtext}
\usepackage{newtxmath}
\usepackage{textcomp}
\usepackage{xcolor}
\usepackage{lipsum}
\usepackage{nicefrac}

\title{Meta-Analysis with JASP, Part II: Bayesian Approaches}

\author{ \href{https://orcid.org/0000-0002-0018-5573}{\includegraphics[scale=0.06]{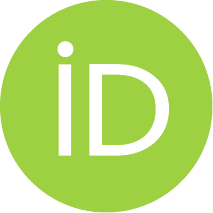}\hspace{1mm}František Bartoš} \\
	Department of Psychological Methods\\
	University of Amsterdam\\
	\texttt{f.bartos96@gmail.com} \\
	\And 
        \href{https://orcid.org/0000-0003-1596-1034}{\includegraphics[scale=0.06]{orcid.pdf}\hspace{1mm}Eric-Jan Wagenmakers} \\
	Department of Psychological Methods\\
	University of Amsterdam
}

\hypersetup{
 pdftitle    = {Meta-Analysis with JASP, Part II: Bayesian Approaches},
 pdfsubject  = {stat.ME},
 pdfauthor   = {František Bartoš, Eric-Jan Wagenmakers},
 pdfkeywords = {Open-Source Software, JASP, Meta-Analysis},
}

\begin{document}
\maketitle

\begin{abstract}
Bayesian inference is on the rise, partly because it allows researchers to quantify parameter uncertainty, evaluate evidence for competing hypotheses, incorporate model ambiguity, and seamlessly update knowledge as information accumulates. All of these advantages apply to the meta-analytic settings; however, advanced Bayesian meta-analytic methodology is often restricted to researchers with programming experience. In order to make these tools available to a wider audience, we implemented state-of-the-art Bayesian meta-analysis methods in the Meta-Analysis module of JASP, a free and open-source statistical software package (\url{https://jasp-stats.org/}). The module allows researchers to conduct Bayesian estimation, hypothesis testing, and model averaging with models such as meta-regression, multilevel meta-analysis, and publication bias adjusted meta-analysis. Results can be interpreted using forest plots, bubble plots, and estimated marginal means. This manuscript provides an overview of the Bayesian meta-analysis tools available in JASP and demonstrates how the software enables researchers of all technical backgrounds to perform advanced Bayesian meta-analysis.
\end{abstract}


\section{Introduction}

Bayesian inference provides a coherent and flexible framework for quantifying uncertainty in parameter estimation \citep{bayes1763problem, gelman1995bayesian, laplace1774memoir, lindley2006understanding}, evaluating evidence for competing hypotheses \citep{jeffreys1935some, wagenmakers2007practical, keysers2020using}, incorporating model uncertainty through Bayesian model averaging \citep{wrinch1921on, leamer1978specification, hoeting1999bayesian}, and updating knowledge as information accumulates \citep{berger1988likelihood, rouder2014optional}. These advantages are particularly pronounced for meta-analysis where studies are often limited in number and may slowly become available over time. Bayesian methods also enable researchers to incorporate prior knowledge, which leads to more stable and reliable estimates of between-study heterogeneity \citep{williams2018bayesian, higgins2009re, chung2013avoiding, rhodes2016implementing, higgins1996borrowing}. Bayes factors \citep{kass1995bayes, jeffreys1961theory} allow informed hypothesis tests that distinguish between \emph{absence of evidence} (i.e., the data are not diagnostic) and \emph{evidence of absence} (i.e., the data provide positive support for the null hypothesis)---both for the treatment effect, for heterogeneity, for moderation, and for publication bias \citep{gronau2021primer, bartos2021bayesian, bartos2025robust, maier2020robust}. Finally, Bayesian model averaging \citep{hinne2019conceptual, hoeting1999bayesian, leamer1978specification} allows researchers to account for uncertainty about the data-generating process \citep{gronau2021primer, bartos2021bayesian, maier2020robust, bartos2021no, bartos2025robust}.

Despite these theoretical and practical benefits, advanced Bayesian meta-analytic methods remain inaccessible to many applied researchers without programming skills. Most implementations are available only in the \texttt{R} programming language \citep{R}, through packages such as \texttt{bayesmeta} \citep{bayesmeta}, \texttt{metaBMA} \citep{metaBMA}, \texttt{RoBMA} \citep{RoBMA}, \texttt{pema} \citep{pema}, \texttt{BFpack} \citep{BFpack}, and \texttt{brms} \citep{brms}. The use of these tools requires coding expertise, which prevents many researchers from adopting Bayesian approaches and forces them to rely on simpler software or to forgo Bayesian analyses altogether.

In order to make advanced Bayesian meta-analytic methods available to a wide range of researchers and students, we recently extended the Meta-Analysis module in JASP, a free and open-source statistical software with a user-friendly graphical interface \citep{JASP95}. The module implements state-of-the-art Bayesian meta-analytic functionality through the \texttt{RoBMA} \texttt{R} package \citep{RoBMA}. This manuscript describes the available Bayesian methods and complements the classical counterpart \citep{bartos2025classical}. Our focus is on statistical methodology rather than practical recommendations for specific applications (see the cited references for guidance). Also see \citet{wagenmakers2023accessible} and \citet{ly2021bayesian} for more details about JASP and \citet{bartos2025classical} for more details about the Meta-Analysis module.

In the sections that follow, we first discuss the specification of the prior distribution. We then demonstrate the Bayesian methods implemented in the JASP Meta-Analysis module using three examples. Example~1 features course instructor ratings and student achievement and illustrates Bayesian parameter estimation, prior and posterior distribution visualization, Markov chain Monte Carlo (MCMC) diagnostics, Bayesian hypothesis testing, and Bayesian model averaging; Example~2 features nitrogen dioxide exposure and respiratory illness in children and demonstrates Bayesian model-averaged meta-regression, estimated marginal means, bubble plot, forest plot, and prior sensitivity analysis; Example~3 features modified school calendars and student achievement and showcases Bayesian multilevel meta-analysis and publication bias adjustments with robust Bayesian meta-analysis. Alongside the examples, we highlight additional options available in the software. Annotated analysis files are available at \url{https://osf.io/7ud8v/}. JASP can be downloaded from \url{https://jasp-stats.org/download/} and the source code is available at \url{https://github.com/jasp-stats/}.

Throughout the manuscript, we assume readers are at least somewhat familiar with JASP, have enabled the Meta-Analysis module, and can load the appropriate datasets. If not, we recommend first consulting the ``Introduction to JASP'' section of \cite{bartos2025classical}. For recent theoretical introductions to Bayesian meta-analysis and Bayesian model-averaged meta-analysis, see \cite{grant2025bayesian}, \cite{mulder2024bayesian}, \cite{gronau2021primer}, and \cite{berkhout2021tutorial} (which accompanies a deprecated version of the module).

\section{Specifying Prior Distributions for Meta-Analyses}

Prior distributions allow researchers to incorporate prior knowledge, regularize parameter estimates, and specify informed hypothesis tests (e.g., \citealp{gronau2021primer, berkhout2021tutorial, mulder2024bayesian} for meta-analysis–specific guidance). Because prior distributions partly define the meta-analytic model, they constitute an essential part of any Bayesian meta-analysis. The JASP Meta-Analysis module provides reasonable default prior distributions for estimation, testing, and model averaging across many standardized effect sizes via the `Effect size measure' option (cf. the top-right section of the left input panel in Figure~\ref{fig:random-effects}). At the same time, the module gives experienced analysts full flexibility to tailor prior distributions to their specific needs. In this section, we describe the default settings, discuss additional options for specifying prior distributions, and provide some general considerations.

The specification of prior distributions is mainly controlled by the following options: the `Effect size measure' dropdown (underneath the `Effect Size' and `Effect Size Standard Error' variable inputs; cf. Figure~\ref{fig:random-effects}), the `Effect size and heterogeneity', `Scale', `Publication bias adjustment' options underneath `Prior Distributions' (cf. the middle-right section of the left input panels of Figures~\ref{fig:random-effects} and~\ref{fig:robma}), and the `Prior Distributions (Custom)' section (grayed out in Figure~\ref{fig:random-effects}). The specified prior distributions can be inspected through the `Show model specification' option.  

\subsection{Effect Size and Heterogeneity}

When the `Effect size and heterogeneity' dropdown option is set to `Default', the specification of the `Effect size measure' invokes the appropriate default prior distribution for the model parameters that corresponds to the scale of the effect size metric (e.g., log odds ratios have a scale that is approximately $1.81$ times larger than that of standardized mean differences, \citealp{borenstein2009introduction}). For standardized mean differences (SMDs, e.g., Cohen’s $d$ or Hedges’ $g$), the default prior distribution for the mean effect size $\mu$ is a standard normal distribution, Normal(mean = 0, sd = 1), and the prior distribution for the between-study heterogeneity $\tau$ is an Inverse-Gamma(shape = 1, scale = 0.15) distribution \citep{vanerp2017estimates}, following the defaults in the \texttt{RoBMA} \texttt{R} package \citep{RoBMA}. If log odds ratios (logOR) or Fisher’s $z$-transformed correlations are specified, the SMD prior distributions are automatically rescaled following the transformations in \citet{borenstein2009introduction}. These defaults have been tested in simulation studies (e.g., \citealp{maier2020robust, bartos2021no}) and work well for estimation, hypothesis testing, and Bayesian model averaging across a wide range of scenarios. Other effect size measures currently do not have default prior distributions. The `Scale' option rescales the defaults to control informativeness: values larger than one make prior distributions less informative, values smaller than one make them more informative.

The `Medicine' option of the `Effect size and heterogeneity' dropdown specifies empirical prior distributions based on previous meta-analyses from the Cochrane Database of Systematic Reviews (CDSR). These prior distributions are available for SMD \citep{bartos2021bayesian}, logOR, log risk ratios (logRR), risk differences (RD), and log hazard ratios (logHR, \citealp{bartos2023empirical}). These prior distributions have slight regularizing properties and are also suitable for parameter estimation, hypothesis testing, as well as Bayesian model averaging in a wide variety of settings. At present, only the medical field has such pre-specified prior distributions.

The `Custom' option of the `Effect size and heterogeneity' dropdown enables a fully flexible specification of prior distribution. In the `Prior Distributions (Custom)' section, users can assign (mixtures of) prior distributions to each parameter under both null and alternative hypotheses (when `Bayesian model averaging' is enabled for a given parameter). This option allows advanced users to specify prior distributions based on published recommendations (e.g., \citealp{pullenayegum2011informed, turner2012predicting, turner2015predictive, lilienthal2023bayesian, gunhan2020random, rover2023using}) or through prior elicitation methods \citep{johnson2010methods, chaloner1996elicitation, ohagan2006uncertain, mikkola2021prior}.

\subsection{Meta-Regression}

Meta-regression is parameterized following the description in \citep{bartos2025robust}. By default, categorical (factor) covariates are transformed into scaled orthonormal contrasts, which guarantees exchangeability of factor levels (i.e., there is no ``default'' level and the prior distribution is specified on differences from the grand mean for each factor level). Continuous covariates are standardized to achieve scale invariance of the prior distributions. These settings result in the meta-regression intercept corresponding to the adjusted effect size (i.e., effect size estimate averaged over the levels of all meta-regression terms); if a meta-regression is specified, the prior distribution for the `Effect size' is assigned to the adjusted effect size (interpretation is further discussed in Example~2).

Under the `Default' settings of the `Effect size and heterogeneity' option, the meta-regression coefficients are set to one-quarter of the default effect size scale as described and tested in \citep{bartos2025robust}. Since the CDSR-specific prior distributions for the `Medicine' `Effect size and heterogeneity' option are more narrow than the `Default' prior distributions, the corresponding prior distributions for meta-regression coefficients are set to $\nicefrac{1}{2}$ of the effect size scale. The `Custom' settings of the `Effect size and heterogeneity' option allow users to specify various prior distributions (and different contrasts) for the meta-regression terms.
 
\subsection{Publication Bias Adjustment}

The `Publication bias adjustment' option (only available in the `Robust Bayesian Meta-Analysis' menu; see Figure~\ref{fig:robma}) lets users choose among different adjustment models. The default `RoBMA-PSMA' option averages across six weight functions and PET-PEESE models as described in \citet{bartos2021no}. Alternatives include `RoBMA-PP' (PET-PEESE, \citealp{stanley2014meta}, adjustment only) and `RoBMA-2w' (the original two–weight function, \citealp{vevea1995general}, approach, \citealp{maier2020robust}). The `Custom' settings allows users to either modify the existing settings (e.g., modifying the selection model cut-points, changing prior distributions for the publication bias adjustment parameters, and adjusting their prior odds) or specify an entirely different (model-averaged) publication bias adjustment based on either step weight functions \citep{vevea1995general}, PET, or PEESE style models \citep{stanley2014meta}.

\subsection{General Considerations}

When specifying custom prior distributions, it is important to recognize the inherent consequences for model behavior and how these consequences might be different for parameter estimation and hypothesis testing/model averaging. On the one hand, setting a very wide ``uninformative'' prior distribution is often considered a conservative choice for parameter estimation; however, ``uninformative'' prior distributions might lead to hypothesis tests heavily biased in favor of the null hypothesis. This bias is a consequence of the resulting overly wide prior predictive distribution under the alternative hypothesis---as a wide prior distribution allows for many possible outcomes, but each must be deemed extremely unlikely. Somewhat paradoxically, in the case of model averaging, the resulting model-averaged estimates will be aggressively shrunk towards the values under the null hypothesis, as the alternative hypothesis receives little posterior model probability. On the other hand, setting a very narrow ``informative'' prior distribution is considered a good practice for obtaining diagnostic tests in hypothesis testing; however, ``informative'' prior distributions might lead to biased posterior estimates as the data may contain much less information than is contained in the prior distributions. 

We generally recommend that the prior distributions are explicitly indicated in any analysis report (see also \citealp{vandoorn2020jasp}) that the robustness to different plausible prior distributions is explored and assessed, and that the prior distributions are carefully thought-through \emph{before} the data are inspected; the latter practice can be enforced by preregistering the analysis plan \citep{hardwicke2023reducing}.  

\section{Example 1: Course Instructor Ratings and Student Achievement}

The first example features the relationship between course instructor ratings and student achievement \citep{cohen1981student}, based on a dataset that comes bundled with the \texttt{metadat} \texttt{R} package \citep{metadat}. The dataset contains correlation coefficients (and sample sizes) between mean instructor rating and mean student achievement score (e.g., final exam grades) from multi-section courses (i.e., courses with sections taught by different instructors but with a common student achievement measure) from twenty studies. Before beginning the analysis, we compute Fisher's $z$ transformation of the correlation coefficients and their standard errors in the `Effect Size Computation' menu. This is done by selecting the `Variable association' option in `Design', `Quantitative' option in `Measurement`, and `ZCOR' option in `Effect size' dropdowns (see \citealp{bartos2025classical} for details on the `Effect Size Calculation' menu).  

\subsection{Parameter Estimation}

\begin{figure}
    \centering
    \includegraphics[width=1\linewidth]{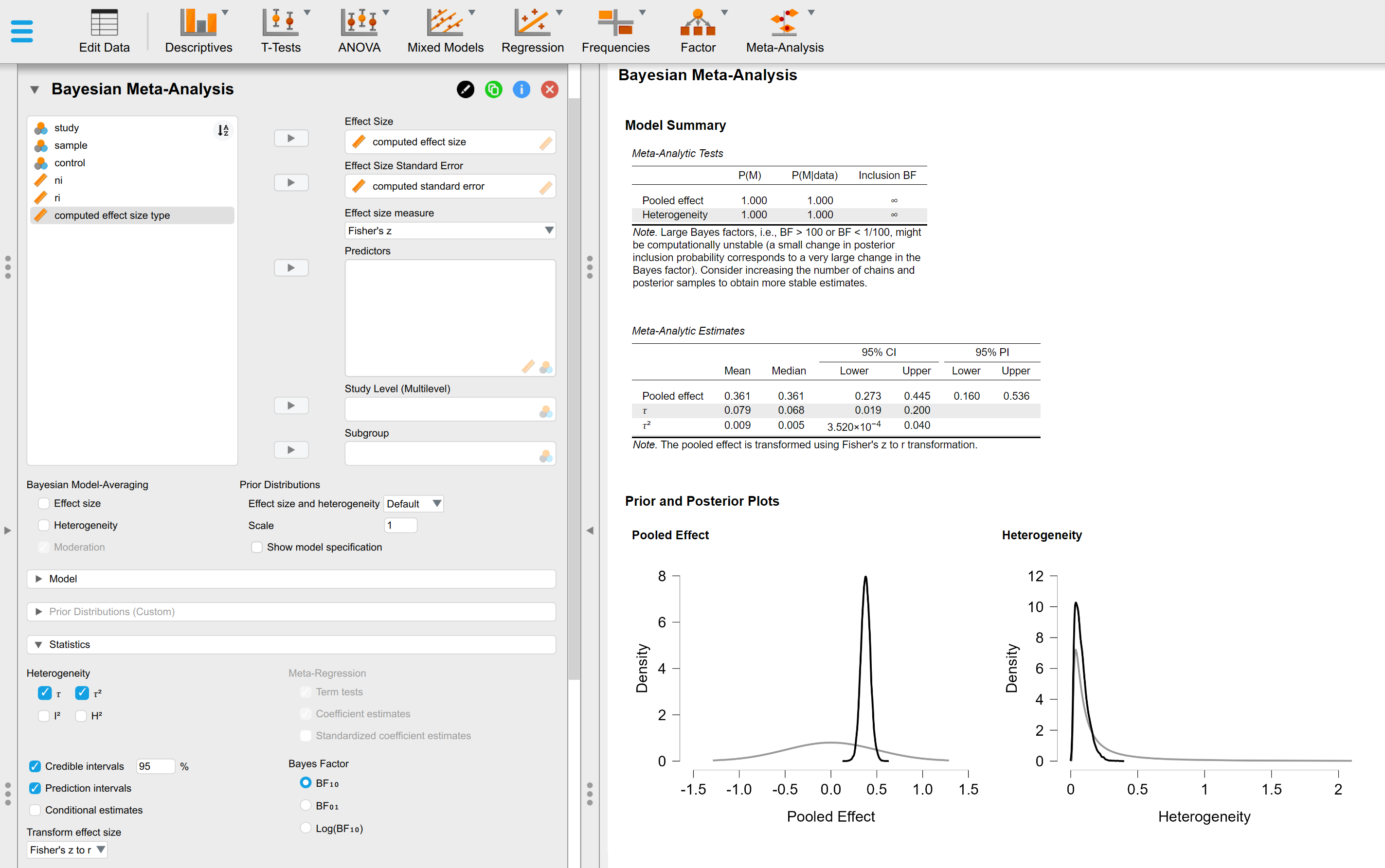}
    \caption{Bayesian Random-Effects Meta-Analysis With Prior and Posterior Distributions}
    \label{fig:random-effects}
\end{figure}

We start with a simple Bayesian random-effects meta-analysis. In the Meta-Analysis module under the `Bayesian' heading, we open the `Meta-Analysis' menu and specify the computed `Effect Size' and `Effect Size Standard Error' variable inputs. The `Effect size measure' dropdown needs to be set to `Fisher's z' so that the default prior distributions are defined on the Fisher's $z$ scale. In the `Bayesian Model Averaging' subsection of the `Model' section, the `Effect size' and `Heterogeneity' options have to be unselected. This disables model averaging across models that assume the absence of the effect and the absence of heterogeneity. Consequently, the analysis setting corresponds to the Bayesian random-effects model assuming the presence of both the effect and heterogeneity (see the upper-left of Figure~\ref{fig:random-effects}). To report the pooled effect on the correlation scale, we set the `Transform effect size' dropdown to the `Fisher's z to r' option in the `Statistics' section. For reproducibility, the `Set seed' option in the `Advanced' section is set to 1 (as well as in all subsequent analyses). Note that exact replication across machines is not guaranteed because operating systems may differ in numerical precision and implementations of JAGS's \citep{JAGS} random number generators.

The `Model Summary' section in the top-right output panel of Figure~\ref{fig:random-effects} shows the main results. Because only a single model is fitted, the `Meta-Analytic Tests' table can be ignored (the prior model probability for both `Pooled effect' and `Heterogeneity' equals 1, which yields a posterior identical to the prior model specification for inclusion and an undefined/infinite inclusion Bayes factor). The `Meta-Analytic Estimates' table summarizes the pooled effect size estimate with central credible and prediction intervals already transformed to the correlation scale. We find a pooled effect $\rho = 0.36$, 95\% central credible interval (CI) [$0.27, 0.45$] indicating a moderate relationship between instructor rating and student achievement, with only small between-study heterogeneity $\tau_\text{Fisher's z} = 0.08$, 95\% CI [$0.02, 0.20$]. Additional heterogeneity statistics, such as I\textsuperscript{2}, are available under the `Heterogeneity' subsection of the `Statistics' section. 

The `Advanced' section provides further control over model fitting. Here we can increase the number of MCMC chains, adaptation, burnin, and sampling iterations \citep{gelman1995bayesian, spiegelhalter2003WinBUGS}. Alternatively, the `Autofit' option can be set up so that MCMC chains are run until a prespecified MCMC convergence criterion is reached or the maximum fitting time is reached. The `Show RoBMA R code' option generates the \texttt{R} code for the specified model, which is helpful when easing into \texttt{R}.

\subsection{Prior and Posterior Distribution Plot}

To visualize the prior and posterior distributions shown in the lower-right panel of Figure~\ref{fig:random-effects}, select the `Effect size' and `Heterogeneity' options in the `Prior and Posterior Plots' section. The `Include prior distribution' option includes the gray density of the prior distribution alongside the black density of the posterior distribution. Note that these plots are not currently transformed by the `Transform effect size' option.

Under hypothesis testing and Bayesian model averaging the prior distribution becomes a mixture of a `spike' at 0 representing the null hypothesis, drawn as an upward facing arrow with a secondary probability axis, and a continuous density representing the prior distribution under the alternative hypothesis (see the lower-right panel of Figure~\ref{fig:bma}). In more complex models, the posterior distribution may be multimodal, reflecting the large model uncertainty; had we entertained only a single model, the posterior distribution would have occupied just one of these modes. A simpler view is available by setting the `Type' option to `Conditional' settings, which shows only the posterior distribution for a parameter (pooled effect or heterogeneity) based on the alternative hypothesis that assumes the presence of the selected parameter. For the simple random-effects meta-analysis (the current example), this setting does not change the figure since the model already assumes both the effect and heterogeneity are present.  

\subsection{MCMC Diagnostics}

\begin{figure}
    \centering
    \includegraphics[width=0.5\linewidth]{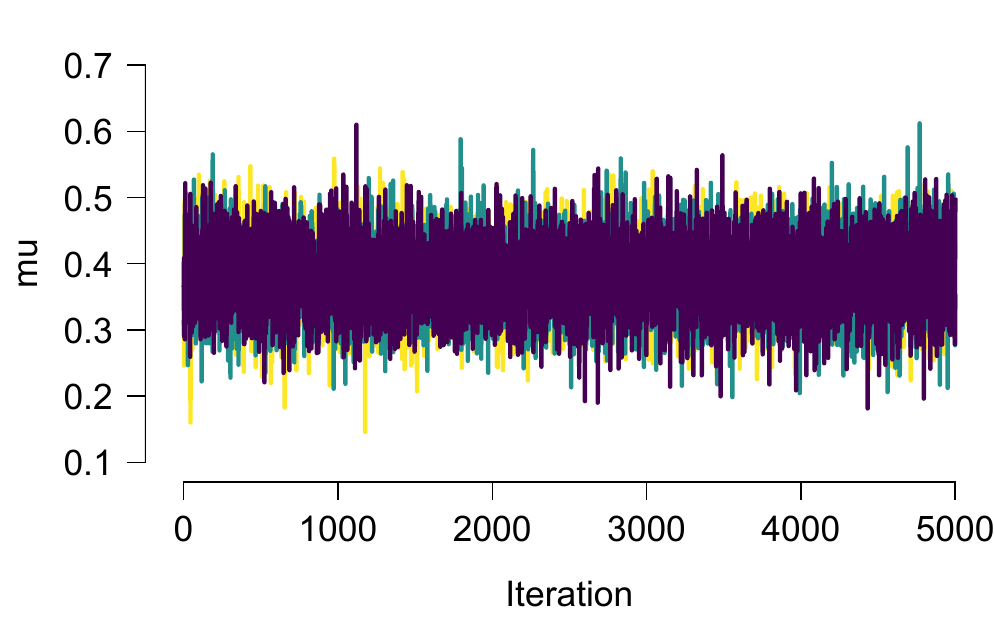}
    \caption{MCMC Diagnostics Trace Plot of the Pooled Effect Created in JASP}
    \label{fig:trace-mu}
\end{figure}

MCMC convergence can be assessed in the `MCMC Diagnostics' section. The `Overview table' reports MCMC error, relative MCMC error, effective sample size (ESS), and $\hat{R}$ \citep{gelman1992inference, van2018simple, mcelreath2020statistical}. Poor MCMC diagnostics are automatically reported in the main `Model Summary' output section. For visual checks, we first select a parameter by means of the corresponding option in the `Plot' subsection  (e.g., `Effect size'), and then select the desired diagnostic type, such as the `Trace' option in the `Type' subsection. Figure~\ref{fig:trace-mu} shows the resulting trace plot of the pooled effect.

Under hypothesis testing and Bayesian model averaging, the interpretation changes in parallel to the prior and posterior distribution plots. Since the hypothesis testing and Bayesian model averaging is implemented via the product space method \citep[e.g.,][]{lodewyckx2011tutorial}, multimodal posterior distributions do not indicate poor chain mixing as long as the individual chains move between the modes. A concerning case would be if different chains were stuck in different modes. Likewise, because models that assume the absence of the effect or of heterogeneity are part of the model space, chains may appear ``stuck'' at the null hypothesis value; this is not problematic unless it occurs for only a single chain.

\subsection{Hypothesis Testing}

The presence versus absence of the pooled effect can be tested by enabling the `Effect size' option in the `Bayesian Model Averaging' subsection. This option selection specifies two models: a Bayesian random-effects meta-analysis that assumes the presence of the effect (i.e., the alternative hypothesis: $\mathcal{H}_1$) and a Bayesian random-effects meta-analysis that assumes the absence of the effect (i.e., the null hypothesis: $\mathcal{H}_0$). The models are estimated jointly via the product space method \cite[e.g.,][]{lodewyckx2011tutorial} implemented in \texttt{RoBMA}. 

The (inclusion) Bayes factor comparing the relative prior predictive performance of the hypotheses is reported in the `Meta-Analytic Tests' table. In our example, the prior model probability `P(M)' of 0.5 shifts to a posterior model probability `P(M|data)' near 1. The inclusion Bayes factor exceeds numerical precision and is reported as $\infty$ (i.e., extreme evidence for the effect). A note below the table warns us that Bayes factors larger than 100 may be unstable. This reflects a limitation of the product space method: distinguishing among very large Bayes factors requires many posterior samples. Although a posterior probability of 0.999 versus 0.998 is practically identical, the corresponding inclusion Bayes factors (999 versus 499) present a seemingly significant numerical difference. Still, the qualitative conclusion---overwhelming evidence for the effect---remains unchanged. Higher precision can be achieved by increasing the number of chains or iterations if needed.  

The `Meta-Analytic Estimates' table in the main output shows the model-averaged estimates. Estimates conditional on the presence of a given parameter are available via the `Conditional estimates' option in the `Statistics' section. In our example, the model-averaged and conditional estimates coincide because the random-effects model that assumes the presence of the effect receives all posterior probability.

\subsection{Model Averaging}

\begin{figure}
    \centering
    \includegraphics[width=1\linewidth]{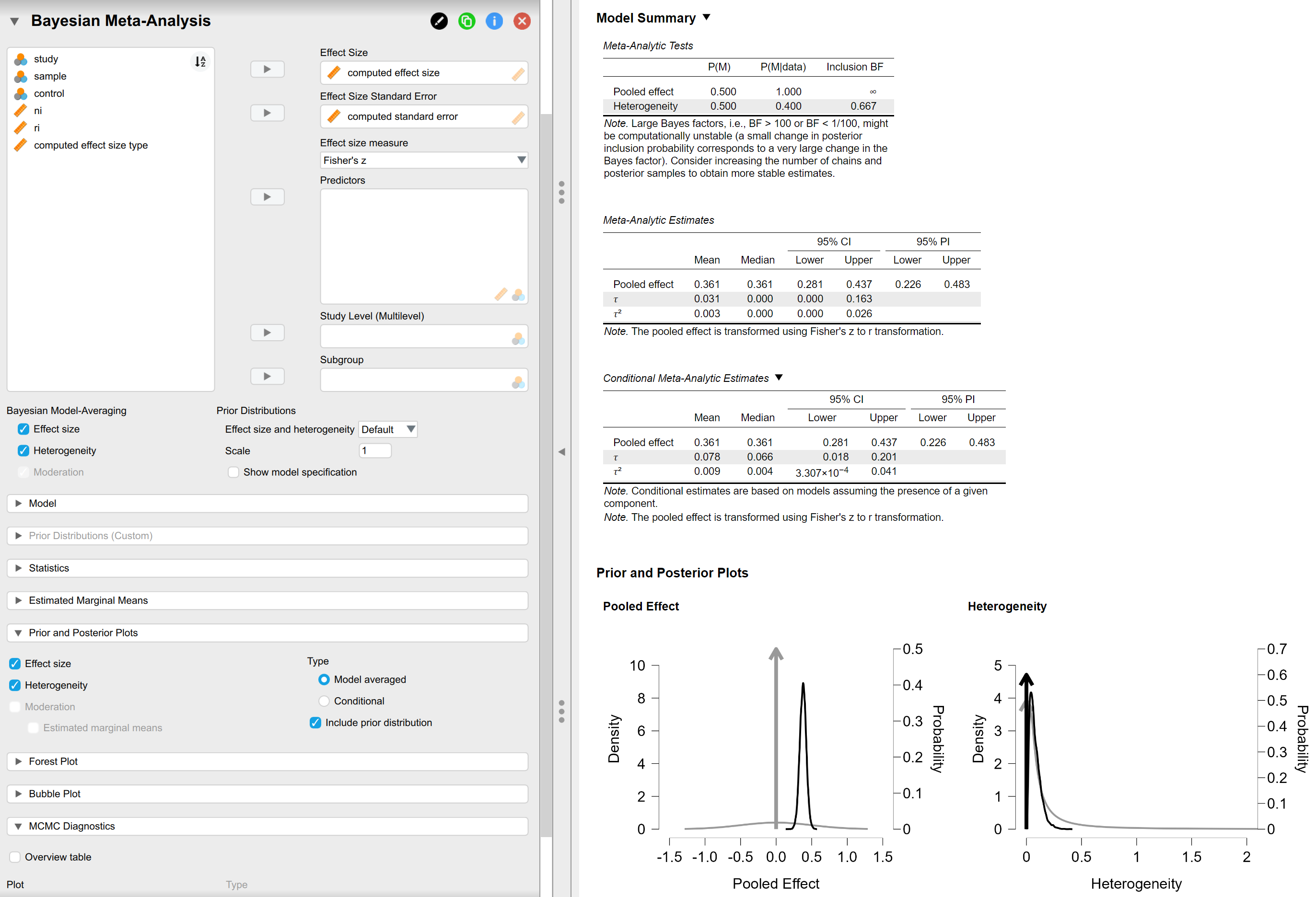}
    \caption{Bayesian Model-Averaged Meta-Analysis With Prior and Posterior Distributions}
    \label{fig:bma}
\end{figure}

The previous specification does not yet account for uncertainty about the presence of between-study heterogeneity. We therefore revert to the default settings of the analysis by also enabling the `Heterogeneity' option in the `Bayesian Model Averaging' subsection (left panel of Figure~\ref{fig:bma}). The product space now includes $2 \times 2$ models: presence versus absence of the effect crossed with presence versus absence of heterogeneity (see \citealp{gronau2021primer} and \citealp{bartos2021bayesian} for more details). Consequently, the inference about the pooled effect takes into account the uncertainty about the presence of heterogeneity, and the inference about the heterogeneity takes into account the uncertainty about the presence of the effect.

In this example, Bayesian model averaging does not meaningfully change the inference about the pooled effect: the evidence for the presence of the effect and the pooled effect estimate remain effectively the same, with only a slight narrowing of the credible and prediction intervals. For heterogeneity, there is weak evidence against its presence, summarized by an inclusion Bayes factor of 0.667 (equivalently, BF\textsubscript{01} = 1/0.667 = 1.5 for exclusion; this computation is also available via `BF\textsubscript{01}' option in the `Bayes Factor' subsection of the `Statistics' section). Consequently, the model-averaged heterogeneity estimate shrinks toward zero relative to models that assume the presence of heterogeneity.

Selecting the `Conditional estimates' option in the `Statistics' section adds a `Conditional Meta-Analytic Estimates' table (middle-right panel of Figure~\ref{fig:bma}) with the effect size and heterogeneity estimated conditional on the models assuming their presence---that is, without shrinkage toward the null hypotheses values. The conditional $\tau$ estimate closely matches the earlier random-effects model that assumed the presence of heterogeneity. The prior and posterior distributions in the lower-right panel of Figure~\ref{fig:bma} now include the mass allocated to the null hypotheses, visualized by the upward facing arrows (see `Prior and Posterior Distribution Plot').

\section{Example 2: Nitrogen Dioxide Exposure and Respiratory Illness in Children}

The second example explores the statistical relationship between nitrogen dioxide (NO\textsubscript{2}) exposure and respiratory illness in children \citep{hasselblad1992synthesis} using a dataset that also comes bundled with the \texttt{metadat} \texttt{R} package \citep{metadat}. The dataset contains odds ratios (ORs, and their confidence intervals) of NO\textsubscript{2} exposure on occurrence of respiratory illness in children from nine studies, along with three covariates: whether or not the study adjusted for parental smoking, whether or not direct NO\textsubscript{2} measurement was used, and whether or not the analysis adjusted for gender. To proceed with the analysis, we first compute log odds ratios (logOR) and their standard errors using the `Effect Size Computation' menu with the `Reported effect sizes' option in the `Design' dropdown.\footnote{Users can transform ORs and their confidence intervals into logORs and corresponding standard errors directly in the JASP data editor.}  

\begin{figure}
    \centering
    \includegraphics[width=1\linewidth]{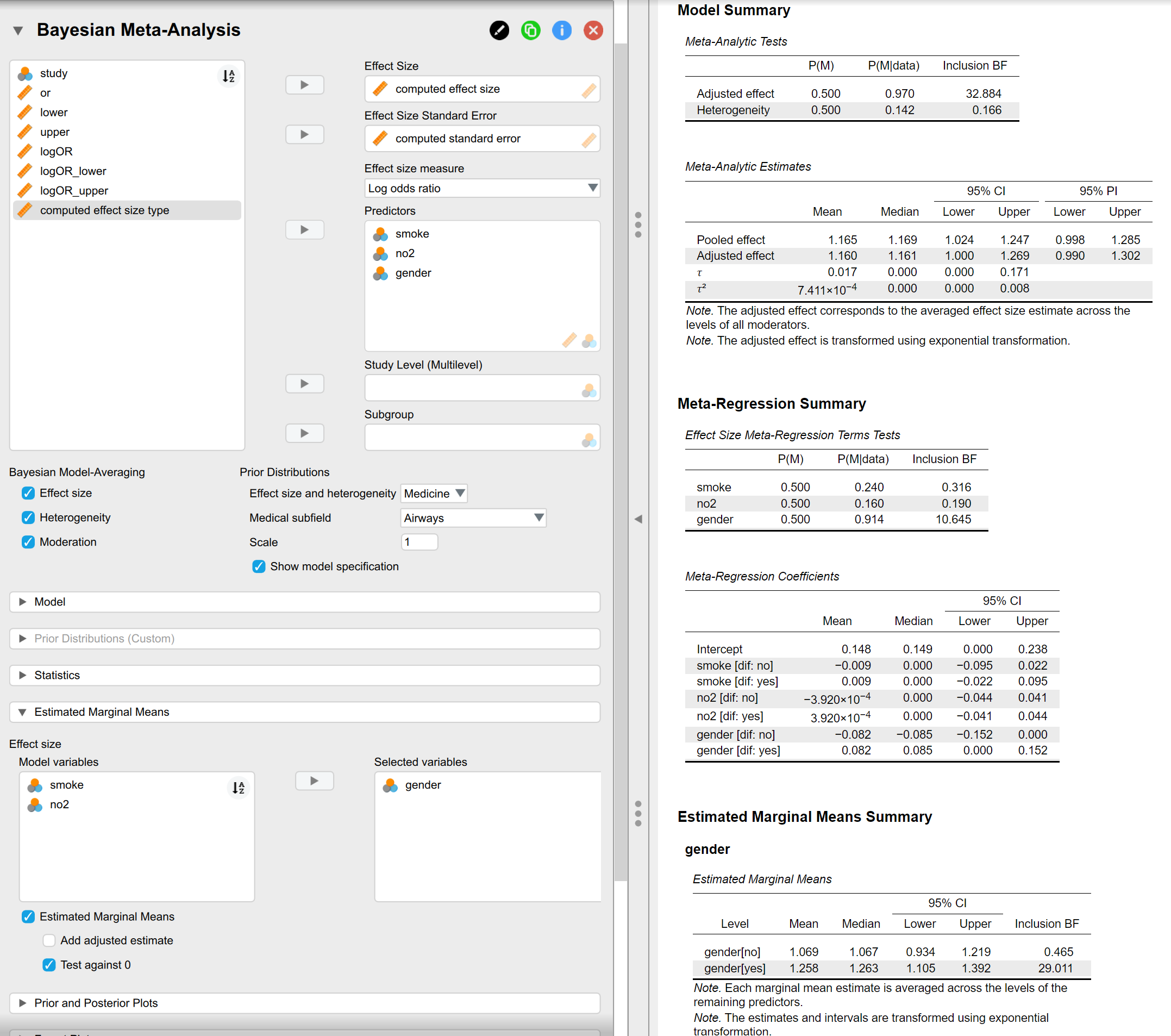}
    \caption{Bayesian Model-Averaged Meta-Regression with Estimated Marginal Means}
    \label{fig:meta-regression}
\end{figure}

\subsection{Meta-Regression}

We analyze the data with a Bayesian model-averaged meta-regression, which allows us to estimate and test for the presence versus absence of the adjusted effect, heterogeneity, and moderation while accounting for uncertainty in all remaining model components. The model is specified by assigning the computed `Effect Size' and `Effect Size Standard Error' variables and by adding the three moderators (`smoke', `no2', `gender') to the `Predictors' input. The meta-regression treats continuous and categorical moderators differently (see \citealp{bartos2025robust} and the `Prior Distributions' section). We therefore verify that categorical moderators are prefaced with the three intersecting circles icon and continuous moderators with the ruler icon; the type can be changed by clicking the icon (input-wise, after assignment) or in the data editor (analysis-wise, before assignment).

Next, the `Effect size measure' dropdown needs to be set to the `Log odds ratio' option in order to obtain the correct scale for the prior distributions. Furthermore, to use the CDSR-based empirical prior distribution for airways-related subfield \citep{bartos2023empirical}, we select the `Medicine' option in the `Effect size and heterogeneity' dropdown and choose the `Airways' option in the `Medical Subfield' dropdown in the `Prior Distributions' subsection. The estimates are reported directly on the OR scale by selecting the `Exponential' option in the `Transform effect size' dropdown in the `Statistics' section. The complete prior specification is available via `Show model specification' in the `Prior Distributions' subsection, and the full input is shown on the right side of Figure~\ref{fig:meta-regression}.

The upper-right panel of Figure~\ref{fig:meta-regression} shows the `Model Summary' with one key change with respect to the simple meta-analysis. The `Meta-Analytic Tests' table now lists the `Adjusted effect' instead of the `Pooled effect', reflecting the default Bayesian meta-regression parameterization that enforces exchangeability across factor levels via standardized mean difference contrasts \citep{bartos2025robust}. The adjusted effect is interpreted as the effect size at the average of the moderator levels (whereas the pooled effect corresponds to the effect size at the average of the moderator values). Consequently, the adjusted and pooled estimates summarized (after exponential transformation) in the `Meta-Analytic Estimates' table may differ if moderator levels are imbalanced. In this example, there is very strong evidence for the adjusted effect and moderate evidence against between-study heterogeneity.

The `Meta-Regression Summary' section in the middle-right panel of Figure~\ref{fig:meta-regression} contains the `Meta-Regression Terms Tests' table, which reports inclusion Bayes factors for each moderator. There is moderate evidence against moderation by parental smoking adjustment and by direct NO\textsubscript{2} measurement, and strong evidence for moderation by gender adjustment status. Individual coefficients appear in the `Meta-Regression Coefficients' table and are interpreted as differences from the adjusted effect for each factor level. When continuous moderators are included, enabling `Standardized coefficient estimates' under the `Meta-Regression' subsection of `Statistics' returns coefficients on the standardized scale implied by the prior distributions. Selecting the `Conditional estimates' option in the `Statistics' section adds a `Conditional Meta-Regression Coefficients' table with estimates conditional on the inclusion of each component.

\begin{figure}
    \centering
    \includegraphics[width=0.5\linewidth]{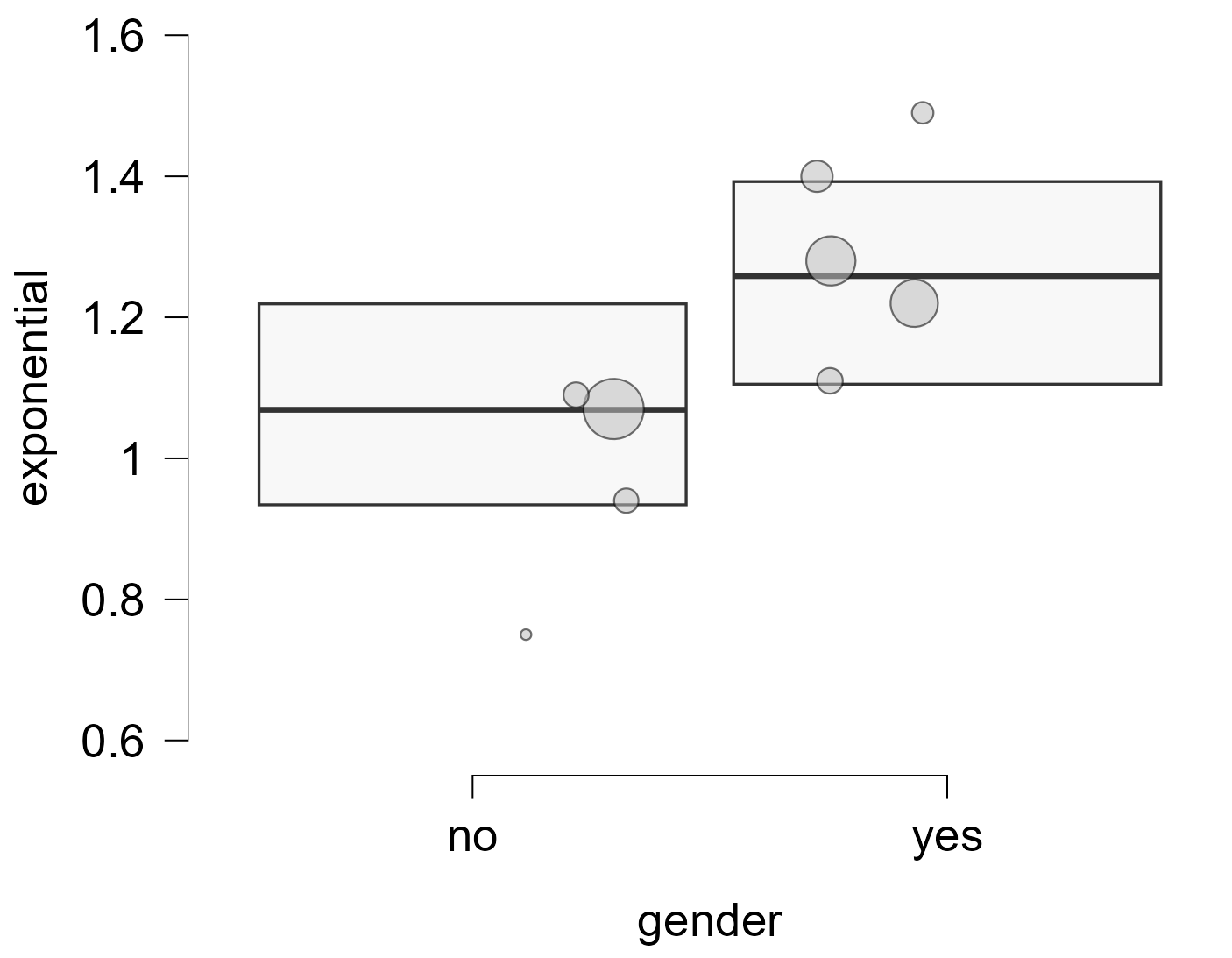}
    \caption{Bubble Plot of Categorical Moderator Created in JASP}
    \label{fig:bubble}
\end{figure}

\subsection{Estimated Marginal Means}

Interpretation of the meta-regression can be further aided by the `Estimated Marginal Means' section. In our example, we examine the gender adjustment moderator. We add `gender' to the `Selected Variables' input in order to obtain estimated marginal means (EMMs): predicted effects for each level of the selected term averaged over the levels of remaining terms. We further select the `Test against 0' option to test the EMM against zero using a Savage-Dickey density ratio test from the implied model prior distribution (see \citealp{bartos2025robust} for details).

The `Estimated Marginal Means Summary' on the bottom-right panel of Figure~\ref{fig:meta-regression} reports the EMM and Bayes factor tests of the EMM against zero. There is weak evidence against the effect when gender is not adjusted for, whereas the gender-adjusted level shows strong evidence for the effect and an OR of about 1.26 with a 95\% credible interval from 1.11 to 1.40. The EMMs are already transformed to ORs following the option chosen under the `Transform effect size' dropdown.

\subsection{Bubble Plot}
The analysis of estimated marginal means can be supplemented with a bubble plot that overlays the estimated effect and its credible interval with the observed effect sizes. For a categorical predictor, the plot shows a box with the estimated marginal mean (thick line) and credible interval (box border) plus jittered study estimates at each moderator level with size determined by their precision (i.e., inverse variance). Figure~\ref{fig:bubble} is obtained by specifying `gender' as the `Selected Variable' input in the `Bubble Plot' section. For a continuous predictor, the visualization displays the estimated trend along the predictor (see \citealp{bartos2025classical} for an example).

\subsection{Forest Plot}

\begin{figure}
    \centering
    \includegraphics[width=1\linewidth]{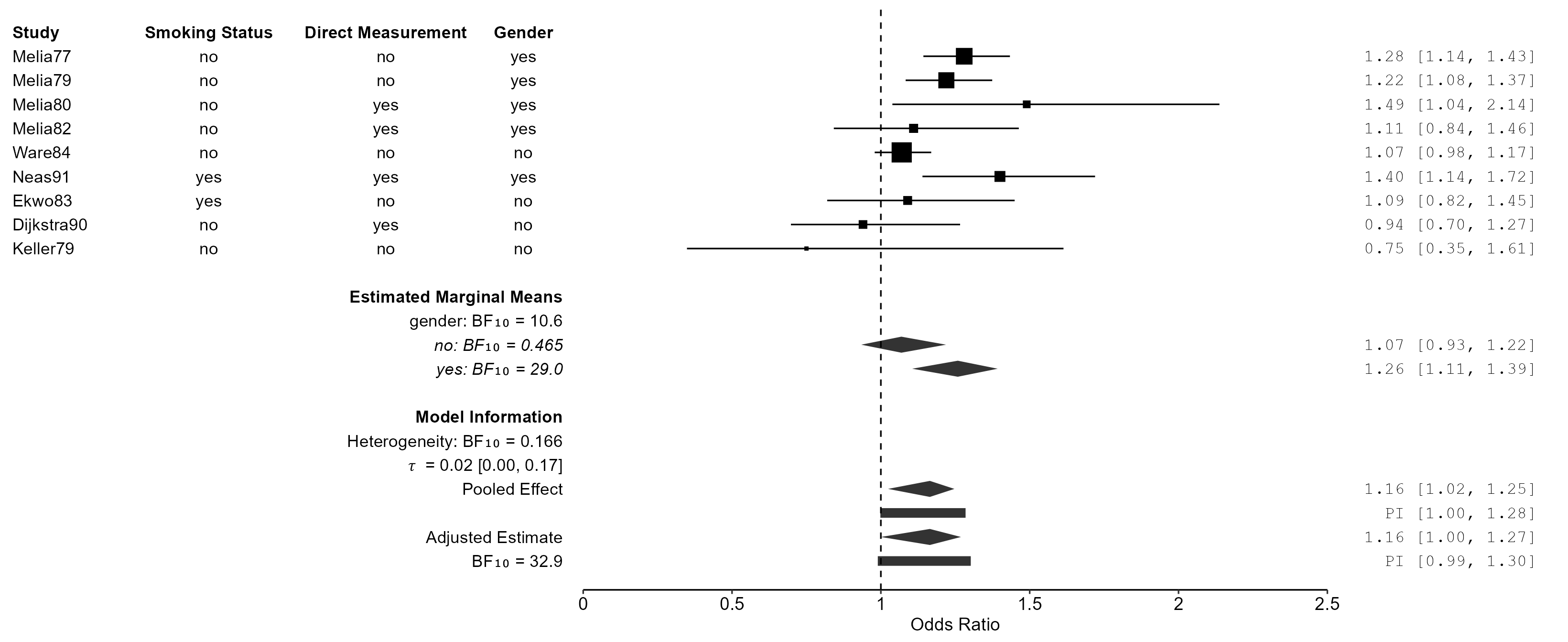}
    \caption{Forest plot created in JASP with study information, estimated marginal means, and model information.}
    \label{fig:forest}
\end{figure}

The forest plot is specified in the `Forest Plot' section. While a complete description of the forest plotting functionality is summarized in \cite{bartos2025classical}, Figure~\ref{fig:forest} shows a forest plot created in JASP that includes the study information section, the estimated marginal means section, and the model information section.

\subsection{Prior Sensitivity Analysis}

A prior sensitivity analysis \citep[see e.g., ][]{box1962further, berger1982robust, berger1990robust} is easily obtained by duplicating the analysis via the `Duplicate this analysis' button (green circular icon with a ``copy'' symbol at the top of the analysis input), specifying an alternative prior distribution, and then comparing the results. In our example, we switch from the subfield-specific `Airways' (`Medicine') prior distribution to the CDSR's `General' empirical prior distribution. The `Model Specification' output highlights that the scale of the prior distribution for effect size is 60\% smaller for the subfield-specific prior distribution than it is for the general prior distribution. Despite this difference, the qualitative conclusions remain the same (see `Prior Sensitivity Analysis' in the analysis file).

\section{Example 3: Modified School Calendars and Student Achievement}

The third example focuses on the effect of modified school calendars on student achievement \citep{konstantopoulos2011fixed} using a dataset that comes bundled (again) with the \texttt{metadat} \texttt{R} package \citep{metadat}. The dataset contains standardized mean differences and sampling variances comparing the academic achievement of students following a modified (i.e., shorter intermittent breaks) versus a traditional (i.e., long summer break) school calendar from 56 studies. The studies were conducted in schools clustered within districts, which motivates a multilevel analysis that allows us to account for study-level and district-level heterogeneity.

\subsection{Multilevel Analysis}

Multilevel meta-analysis relaxes the independence assumption of simple meta-analytic models. This assumption is often violated when estimates from a single study resemble each other more than estimates from different studies, for instance due to similar manipulations, participant populations, or measures within studies \citep{konstantopoulos2011fixed, nakagawa2012methodological}. Ignoring this dependence can yield overconfident results, with credible intervals that are too narrow and Bayes factors that are overly extreme. Multilevel models address the dependence by estimating heterogeneity at each ``level'' of the analysis (i.e., study level and estimate level).

\begin{figure}
    \centering
    \includegraphics[width=1\linewidth]{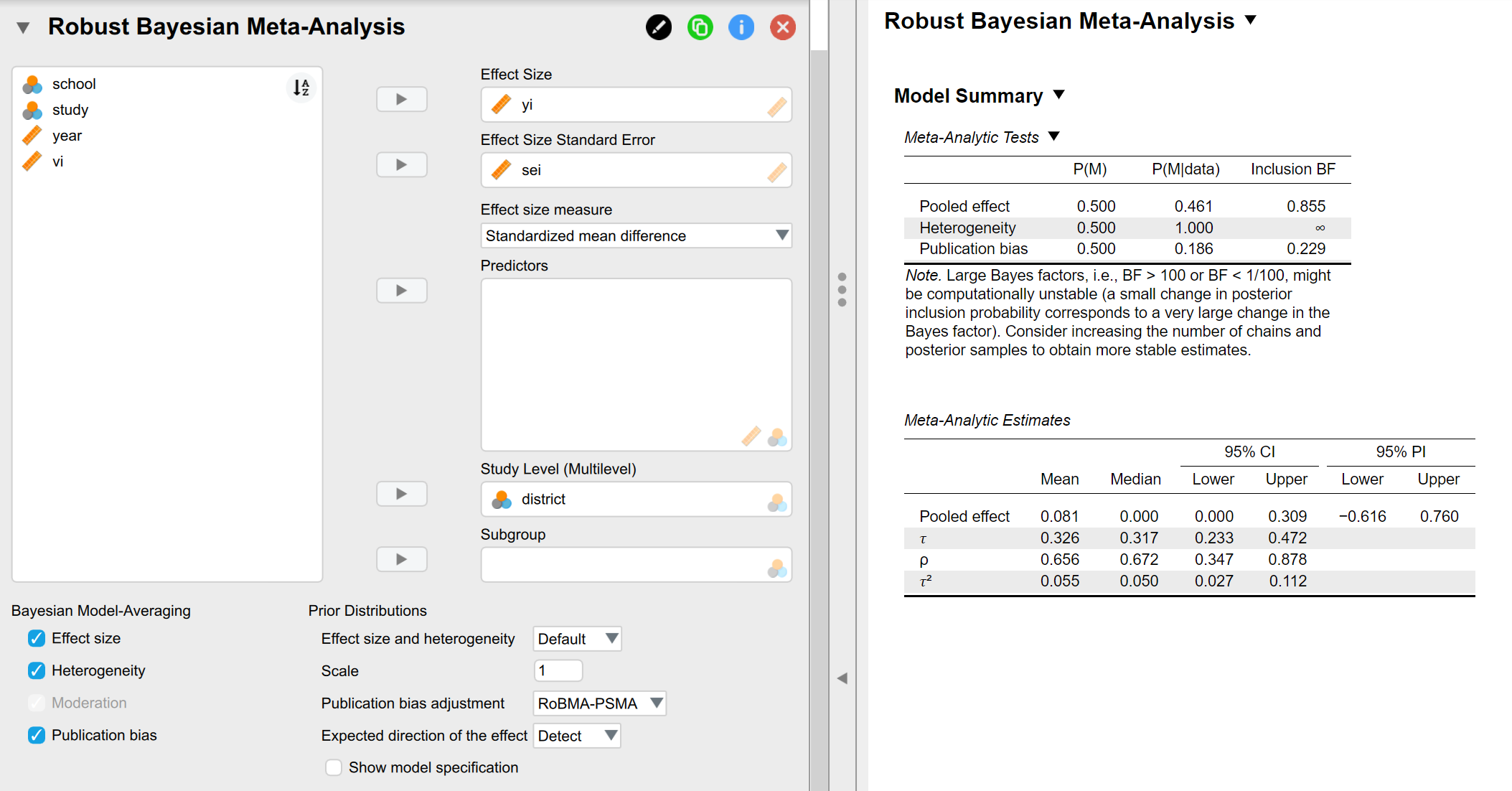}
    \caption{Multilevel Robust Bayesian Meta-Analysis}
    \label{fig:robma}
\end{figure}

The Bayesian multilevel meta-analysis is specified in the `Meta-Analysis' menu. It requires specification of the `Effect Size' and `Effect Size Standard Error' variable inputs. In addition, the `Study Level (Multilevel)' variable input needs to be set in order to define the nesting structure. In this example, the `Study Level (Multilevel)' variable corresponds to the `district', indicating which estimates belong to the same district (usually it will be the study variable indicating which estimates belong to the same study). We keep the default specification in the `Bayesian Model Averaging' subsection to fit a multilevel Bayesian model-averaged meta-analysis \citep{bartos2025robust2}, and we keep the `Default' setting in the `Prior Distributions' subsection for the `Effect size and heterogeneity' dropdown with the `Standardized mean difference' option in the `Effect size measure' dropdown (see the left panel of Figure~\ref{fig:robma} for a matching input setup, mirrored from the `Robust Bayesian Meta-Analysis' specification introduced below).

At first, a note in the `Model Summary' output warns that the model fit did not achieve the desired effective sample size of 500. To resolve this warning, we increase the number of `Sampling' iterations in the `MCMC' subsection of the `Advanced' section to 15{,}000. This adjustment clears the warning. The primary output mirrors the structure familiar from previous examples: the `Meta-Analytic Tests' table reveals that the data provide weak evidence against the effect and extreme evidence for heterogeneity.\footnote{The weak evidence for the effect is not too dissimilar from the just statistically significant result obtainable via the classical multilevel meta-analysis \citep{benjamin2018redefine}.} The `Meta-Analytic Estimates' table is extended with the $\rho$ estimate that represents the variance allocation across levels. We find $\rho = 0.66$, indicating that 66\% of the total variance ($\tau^2$) is attributed to the higher, study-level (i.e., between-district) component. Values of $\rho$ near 0 indicate that most variance lies at the estimate level (estimates are independent); values near 1 indicate that most variance lies at the study level (estimates are dependent).

\subsection{Publication Bias Adjustment}

Finally, we extend the analysis to adjust for publication bias by specifying the equivalent inputs in the `Robust Bayesian Meta-Analysis' menu. The key extension is the `Publication bias' option in the `Bayesian Model Averaging' subsection, which adds publication-bias models to the averaging, and the `Publication bias adjustment' dropdown in the `Prior Distributions' subsection, which selects the specific publication bias adjustment models. The default `RoBMA-PSMA' option applies a mixture of six weight functions \citep{vevea1995general} and PET-PEESE-style models \citep{stanley2014meta}, as summarized in \citet{bartos2021no}. See the left panel of Figure~\ref{fig:robma} for the full input specification. 

The right panel of Figure~\ref{fig:robma} shows the extended `Model Summary' output section. The `Meta-Analytic Tests' table now includes the `Publication bias' row that summarizes the evidence for publication bias. The results indicate moderate evidence against the presence of publication bias, with negligible impact on the evidence for the `Pooled effect' or for `Heterogeneity'. The `Meta-Analytic Estimates' table likewise shows no meaningful difference, suggesting that the multilevel results are robust to publication bias adjustment.

The `Robust Bayesian Meta-Analysis' section also provides additional summaries. The `Statistics' section includes a `Publication bias adjustment' subsection with `Weight function estimates' and `PET-PEESE estimates' tables. The `Priors and Posterior Plots' section allows direct visualization of the estimated weight function and the PET-PEESE regression line. Finally, the `MCMC Diagnostics' section provides the corresponding MCMC assessment.

Custom publication bias adjustments can also be specified under the `Prior Distributions (Custom)' section after selecting the `Custom' option in the `Publication bias adjustment' dropdown. This allows users to fit simple Bayesian selection models \citep{larose1998modeling, smith2001adjustment, givens1995estimating, silliman1997hierarchical}, the Bayesian version of PET and PEESE models \citep{bartos2021no}, or adjust the prior weights of the prespecified components. Publication bias adjustment can be combined with any of the methods discussed earlier, including meta-regression \citep{bartos2021no, bartos2025robust, bartos2025robust2}.

\section{Conclusion}

We have provided an overview of the Bayesian meta-analysis functionality in the JASP Meta-Analysis module. The options and output closely parallel those of the classical meta-analysis implementation in the companion paper by \citet{bartos2025classical}. JASP's point-and-click graphical user interface allows users who are new to Bayesian methods to perform state-of-the-art Bayesian meta-analyses without writing code. Analysts as well as students can therefore focus on applying the most appropriate methods for their research question (and on properly interpreting the results) rather than on constructing and checking complex analysis scripts. In addition, the fully reproducible JASP analysis files can be annotated and shared directly via the Open Science Framework, further promoting reproducibility and transparency.  

Compared to the more mature classical methodology, the Bayesian implementation via the \texttt{RoBMA} \texttt{R} package \citep{RoBMA} is still missing several advanced features. These include location–scale models \citep{viechtbauer2022location}, contrasts of estimated marginal means, and complex multivariate dependency structures \citep{olkin2009stochastically}. At the same time, the Bayesian approach surpasses classical analyses in two key respects: it allows direct quantification of evidence in favor of or against the null hypothesis \citep{wagenmakers2007practical, keysers2020using}, and it naturally incorporates model uncertainty through Bayesian model averaging \citep{hinne2019conceptual, hoeting1999bayesian}. The integration of publication bias adjustment with Bayesian model averaging, multilevel models, and meta-regression further extends the scope of Bayesian methodology.  

In summary, the Bayesian meta-analytic features implemented in JASP make advanced methodology broadly accessible. The free and open-source nature of the software, combined with growing institutional support, ensures that JASP is a sustainable platform for evidence synthesis and well-positioned to adapt to future methodological developments.  


\paragraph{Acknowledgments}
František Bartoš is grateful to the JASP Team for reviewing the Meta-Analysis module pull requests and developing new features necessary to implement the module. 

\paragraph{Funding Statement}
None.

\paragraph{Competing Interests}
František Bartoš and Eric-Jan Wagenmakers declare their involvement in the open-source software package JASP (\url{jasp-stats.org}), a non-commercial, publicly funded effort to make Bayesian statistics accessible to a broader group of researchers and students. In addition, František Bartoš and Eric-Jan Wagenmakers are involved in JASP Services BV, a company that supports organizations who wish to adopt JASP, with a focus on applications in quality control.

\paragraph{Data Availability Statement}
JASP can be downloaded from \url{https://jasp-stats.org/download/} with the source code available at \url{https://github.com/jasp-stats/}. Annotated example analysis files are available at \url{https://osf.io/7ud8v/}.  

\paragraph{Ethical Standards}
The research meets all ethical guidelines, including adherence to the legal requirements of the study country.

\paragraph{Author Contributions}
Conceptualization: F.B. Methodology: F.B. Data curation: F.B. Data visualisation: F.B. Writing original draft: F.B.; E.J.W. All authors approved the final submitted draft.


\bibliographystyle{biometrika}
\bibliography{bib/all.bib, bib/software, bib/unpublished} 

\end{document}